\documentclass[aps,prd,twocolumn]{revtex4}

\usepackage{amsmath,amssymb,amsthm,graphicx,latexsym, epsf}

\newcommand{\be}{\begin{equation}}
\newcommand{\ee}{\end{equation}}
\newcommand{\bea}{\begin{eqnarray}}
\newcommand{\eea}{\end{eqnarray}}
\newcommand{\bml}{\begin{subequations}}
\newcommand{\eml}{\end{subequations}}
\newcommand{\bfig}{\begin{figure}}
\newcommand{\efig}{\end{figure}}

\begin{document}

\title{Reheating and leptogenesis in a SUGRA inspired brane inflation}

\author{Sayantan Choudhury$^{1}$\footnote{Electronic address: {sayanphysicsisi@gmail.com}} ${}^{}$
and Supratik Pal$^{1, 2}$\footnote{Electronic address: {supratik@isical.ac.in}} ${}^{}$}
\affiliation{$^1$Physics and Applied Mathematics Unit, Indian Statistical Institute, 203 B.T. Road, Kolkata 700 108, India \\
$^2$Bethe Center for Theoretical Physics and Physikalisches Institut der Universit\"{a}t Bonn, Nussallee 12, 53115 Bonn, Germany}



\begin{abstract}

We have studied extensively phenomenological implications in
 a specific model of brane inflation  driven by background supergravity \cite{sayan},
 via thermal history of the universe and leptogenesis pertaining
to the particle physics phenomenology of the early
universe. Using the one loop corrected  inflationary potential  we have investigated for
 the analytical expression as well as the numerical estimation for brane reheating temperature  for standard model
 particles. This results in some novel features of reheating from this type of inflation which have serious
implications in the production of heavy Majorana neutrinos needed for leptogenesis through the reheating temperature.
  We have also derived the expressions for the gravitino abundance during reheating
  and radiation dominated era. We have
further estimated different  parameters at the epoch of
 phase transition and revealed their salient features.
At the end we have explicitly given an estimate of the amount of CP violation
through the effective CP phase which is related to baryon asymmetry as well as gravitino dark matter abundance.

\end{abstract}

\maketitle


\section{\bf Introduction}

It is now  well accepted  that the post big bang universe \cite{haw}
 passed through different phases having two-fold
significance -- phenomenological and cosmological. One of the significant phases, namely, reheating \cite{kofman}
 plays the pivotal role in explaining
 production of different particles from inflaton/ vacuum energy. As we look back
in time  reheating was completed within the first second
(and probably much earlier) after the big bang. At that time nucleosynthesis \cite{burles},
or the formation of light nuclei occurred. Particle physicists as well as
 cosmologists have a clear picture of this hot big bang phase because
 ordinary matter and radiation were driving it and also the physical processes that
  characterize it involve terrestrial physics. On the other hand
the mysterious force that drives the inflationary phase is conventionally described
 by a scalar field, named  inflaton which oscillates
near the minimum of its effective potential and produces
elementary particles \cite{palma}. These particles interact with each
other and eventually they come to a state of thermal
equilibrium at some arbitrary temperature T. This process completes
when all the energy of the classical
scalar field transfer to the thermal energy of elementary
particles. Since long theoretical physicists have been investigating reheating as a perturbative phase \cite{Mazumdar}, or one
 in which single inflaton quanta decayed individually into ordinary matter and
 radiation. The recent theoretical studies have shown that in many cases the decay
 occurs through a non-perturbative process \cite{lee}, in which the particles behave in an
 ordered manner. Non-perturbative processes involved at reheating are extremely
 more efficient than the perturbative ones \cite{allah} and often more difficult to investigate
in practice. In short there is
 no existence of a complete theory which explains non-perturbative effects during reheating
 for the total time scale.

     Besides production of gravitinos during perturbative reheating \cite{seto,raga,bolz}  its
 decay plays a significant role in the context of leptogenesis  \cite{pila,van}. More precisely
 two types of gravitinos are produced in this epoch - stable \cite{yana} and unstable \cite{Kohri}.
 Stable ones and decay products of
unstable ones directly or indirectly stimulate the light element
abundances during big bang nucleosynthesis. Most importantly the
 unstable one has important cosmological consequences out of which the major one directly affects the expansion
rate of the universe \cite{Moroi}. In order to explain cosmological consequences at a time by a single physical entity, it
is customary to explain everything in terms of gravitino energy density which is directly
 proportional to the gravitino number density or gravitino
abundance. This gravitino abundance is obtained by
considering gravitino production in the radiation dominated
era following reheating \cite{Choi}. Gravitinos are originated through thermal scattering \cite{bolz,prad} in the
early universe and are usually related to the reheating temperature ($T^{reh}$).
 Particle physics phenomenology usually requires that under instantaneous decay
 approximation \cite{kolb} reheating temperature ($T^{reh}$) is maximum during
  reheating.

 In the present article we have studied extensively reheating phenomenology and leptogenesis
in a typical brane inflation model which was proposed earlier by us \cite{sayan}.
Precisely, the model includes one loop radiative correction in the framework
of local brane version of the supersymmetric theory i.e.
 $N=1,D=4$ SUGRA which is derived from the background $N=2,D=5$ SUGRA in the bulk (for details please refer to \cite{sayan}). In the present article, our prime objective is to investigate for both
 the analytical and numerical expression for brane reheating temperature
 in high energy limit for standard model particles followed by a detailed investigation for
gravitino phenomenology and leptogenesis. Here, and throughout the rest of the article, by
high energy limit implies that the total energy density is very high with respect to the brane tension as mentioned
in our earlier paper \cite{sayan}.
As it will be revealed,   the  scenario is somewhat different in the context of reheating from brane inflation which results in novel features worth studying in details. This has serious implication for the production of the heavy
   Majorana neutrinos needed for leptogenesis \cite{van}.
We further estimate different  parameters related to reheating and leptogenesis at the epoch of
 phase transition \cite{paolo}. Last but not the least we have given an estimate of CP violation which is the indirect evidence of
the baryon asymmetry and connected with gravitino dark matter abundance.


\section{\bf Reheating phenomenology on the brane for $\bf{SU(2)_{L}\bigotimes U(1)_{Y}}$}
\subsection{Model Building from background supergravity}

For systematic development of the formalism, let us 
briefly review from our previous paper \cite{sayan} how one can construct the effective 4D inflationary
potential of our consideration starting from $N=2, D=5$ SUGRA in
the bulk which leads to an effective $N=1, D=4$ SUGRA in the brane. Considering the fifth dimension 
is compactified on the orbifold $S^{1}/Z_{2}$ of
comoving radius R, the  $N=2,D=5$ bulk SUGRA is described by the following action
\be\begin{array}{ll}\label{su1}
 \displaystyle S=\frac{1}{2}\int d^{4}x\int^{+\pi
R}_{-\pi R}dy\sqrt{g_{5}}\left[M^{3}_{5}\left(R_{(5)}-2\Lambda_{5}\right)+L^{(5)}_{SUGRA}\right.\\ \left.~~~~~~~~~~~~~~~~~~~~~~~~~~~~~~~~~~~~~~~~~~~
+\sum^{2}_{i=1}
\delta(y-y_{i})L_{4i}\right].\end{array}\ee Here the sum includes the walls at the orbifold points
$y_{i}=(0,\pi R)$ and 5-dimensional coordinates
$x^{m}=(x^{\alpha},y)$, where $y$ parameterizes the extra
dimension compactified on the closed interval $[-\pi R,+\pi R]$. Written explicitly, the
contribution from bulk SUGRA in the action
\be\begin{array}{ll}\label{sug2}e^{-1}_{(5)}L^{(5)}_{SUGRA}=-\frac{M^{3}_{5}R^{(5)}}{2}+\frac{i}{2}\bar{\Psi}_{i\tilde{m}}
\Gamma^{\tilde{m}\tilde{n}\tilde{q}}\nabla_{\tilde{n}}
\Psi^{i}_{\tilde{q}}-{S}_{IJ}F^{I}_{\tilde{m}\tilde{n}}F^{I\tilde{m}\tilde{n}}\\
~~~~~~~~~~~~~~~~~~~~~~-\frac{1}{2}g_{\alpha\beta}(D_{\tilde{m}}\phi^{\mu})(D^{\tilde{m}}\phi^{\nu})
 + {\rm Fermionic} \\~~~~~~~~~~~~~~~~~~~~~~~~~~~~~~~~~~~~+ {\rm Chern-Simons},\end{array}\ee
 and including the radion fields ($\chi,T,T^{\dag}$)
 the effective brane SUGRA counterpart turns out
to be $\delta(y)L_{4}=-e_{(5)}\Delta(y)\left[(\partial_{\alpha}\phi)^{\dagger}(\partial^{\alpha}\phi)
+i\bar{\chi}\bar{\sigma}^{\alpha}D_{\alpha}\chi\right]$. The Chern-Simons terms can be gauged away
assuming cubic constraints and $Z_2$
symmetry. Further, $S^{1}/Z_{2}$ orbifold setting allows us to express the 4-dimensional part of the action 
(after dimensional reduction) as,
\be\begin{array}{ll}\label{ast8}S=\frac{M^{2}_{PL}}{2}\int d^{4}x \sqrt{g_{4}}\left[R_{(4)}+(\partial_{\alpha}\phi^{\mu})^{\dag}(\partial^{\alpha}\phi_{\mu})-QV_{F}
\right.\\ \left.~~~~~~~~~~~~~~~~~-P\int^{+\pi R}_{-\pi R}dy \frac{4(3e^{2\beta y}
+3\lambda^{2}e^{-2\beta y}-2\lambda)}{R^{2}(e^{\beta y}+\lambda e^{-\beta y})^{5}}\right].\end{array}\ee
where $P=\frac{2M^{3}_{5}\beta b^{6}_{0}}{M^{2}_{PL}R^{5}},~~Q=\frac{C(T,T^{\dag})}{4\pi^{2}R^{2}}$
 and the 4D Planck mass $M_{PL}=\frac{e_{4}}{b_{0}}=\sqrt{\frac{6}{\lambda}e_{(5)}}=\sqrt{8\pi}M=\frac{M^{3}_{5}}{\sqrt{\lambda}}\sqrt{\frac{3}{4\pi}}=1.22\times 10^{19}GeV$.
 Here we have introduced the reduced
4D Planck mass $M=2.43\times 10^{18}GeV$,  5D and 4D charge $e_{5}$ and $e_{4}$,
 5D Planck mass $M_{5}$ and the brane tension $\lambda$ and two constants $\beta$ and $b_{0}$ comes from the metric structure.
 Here $C(T,T^{\dagger})$  represents an arbitrary
function of $T$ and $T^{\dagger}$.
This leads to an effective $N=1, D=4$
SUGRA in the brane with the F-term potential
 \be\begin{array}{lll}\label{vf}
V=V_{F}=\exp\left(\frac{K(\phi,\phi^{\dagger})}{M^{2}}\right)\left[\left(\frac{\partial W}{\partial
\Psi_{\alpha}}+\left(\frac{\partial K}{\partial
\Psi_{\alpha}}\right)\frac{W}{M^{2}}\right)^{\dag}\right.\\ \left.~~~~~~~~~~~~~~ \left(\frac{\partial^{2}K}{\partial\Psi^{\alpha}\partial\Psi^{\dagger}_{\beta}}\right)^{-1}\left(\frac{\partial W}{\partial
\Psi^{\beta}}+\left(\frac{\partial K}{\partial
\Psi^{\beta}}\right)\frac{W}{M^{2}}\right)-3\frac{|W|^{2}}{M^{2}}\right].\end{array}\ee
  Here $\Psi^{\alpha}$ is the chiral superfield and $\phi^{\alpha}$
be the 4D complex scalar field. In this context
the K$\ddot{a}$hler potential is dominated by the leading order
term
i.e. $K= \sum_{\alpha}\phi^{\dagger}_{\alpha}\phi^{\alpha}$.
The superpotential in eqn(\ref{vf}) is given by
$W=\sum^{\infty}_{n=0}D_{n}W_{n}(\phi^{\alpha})$
 with the constraint $D_{0}=1$. Expanding the slowly varying inflaton potential around the value of the inflaton field
along with $Z_{2}$ symmetry the required renormalizable one-loop corrected inflaton potential turns
out to be  
\be\label{opdsa}
V(\phi)=\Delta^{4}\left[1+\left(D_{4}+K_{4}\ln\left(\frac{\phi}{M}\right)\right)
\left(\frac{\phi}{M}\right)^{4}\right],\ee where 
$K_{4}=\frac{9\Delta^{4}C^{2}_{4}}{2\pi^{2}M^{4}}$ and
$D_{4}=C_{4}-\frac{25K_{4}}{12}$ where $C_4$ is negative constant appearing at the tree level.
 Here $\Delta$ represents the energy scale of brane inflation
which can be expressed in terms of the slow roll parameter $\eta_{v}$ explicitly derived in \cite{sayan}. For our model $\Delta\simeq 0.2\times10^{16}GeV$
for the window $ -0.70 < D_{4} <-0.60$.

With this brief review of the construction of the potential we are
 now in a position to investigate for its phenomenological significances.

 From the knowledge of particle physics it is known that during the epoch of reheating inflatons
decay into different particle constituents \cite{kofman,chung} are directly related to the
 trilinear coupling of the inflaton field. There might be a possibility of collision originated
 through quartic coupling and driven by background scalar field. For example here the contribution from the heavy Majorana neutrino comes
from the seesaw Lagrangian ${\cal L}_{Majo}=-h\bar{l}_{L}{\cal H}\psi-\frac{1}{2}{\cal M}\bar{\psi}\psi+h.c.$, where
 $l_{L}$ and ${\cal H}$ are
the lepton and the Higgs doublets, respectively, and ${\cal M}$ is the lepton-number-violating mass
term of the right-handed neutrino. Now using the assumption $ m_{\phi}\gg m_{\sigma}$,$m_{\phi}\gg
m_{\psi}$ the {\it total inflaton decay width} for the positively and negatively charged
$\phi(\phi^{+},\phi^{-})$ scalar fields as well as the fermionic field $\psi$ (Example: For the heavy Majorana neutrinos the decay
process $\psi\rightarrow l_{L}{\cal H}$, $\psi\rightarrow \bar{l}_{L}{\cal H}$ predominates.)
 is given by $\Gamma_{total}\simeq \frac{C^2}{16\pi
m_{\phi}}+\frac{h^{2}m_{\psi}}{4\pi}\sim
\frac{1}{(2\pi)^{3}}\left(\frac{\Delta^{6}}{M^{5}}\right)$ where
the coupling strength
$C\sim m_{\phi}\left(\frac{\Delta^{2}}{M^{2}}\right)$ and $h\sim\left(\frac{\Delta^{2}}{M^{2}}\right)$ and the background
scalar field is $\sigma$.

Now to construct the thermodynamical observable the effective number of particles incorporating relativistic degrees of freedom is defined \cite{turner} as
 $N^{*}=N^{*}_{B}+\frac{7}{8}N^{*}_{F}$,  where $N^{*}_{B}=\sum_{i}N^{*}_{Bi}$ and $N^{*}_{F}=\sum_{j}N^{*}_{Fj}$. Here $N^{*}_{B}$ represents the number of
bosonic degrees of freedom with mass $m_{\phi}\ll T$ and
$N^{*}_{F}$ represents number of fermionic degrees of freedom with
mass $m_{\psi}\ll T$. Here `i' and `j' stand for different
bosonic and fermionic species respectively. For the phenomenological
estimation \cite{ali} $N^{*}\sim 10^{2}-10^{4}$ and for realistic
models $N^{*}\sim 10^{2}-10^{3}$. For convenience let us express reheating
temperature on the brane as
\be\label{ga}\Gamma_{total}=3H(T^{br})=\sqrt{\frac{3\rho(t_{reh})}{M^{2}}
\left[1+\frac{\rho(t_{reh})}{2\lambda}\right]},\ee
where $H(T^{br})$ and $\rho(t_{reh})$ be the Hubble parameter and energy density during reheating respectively.
It is worth mentioning that the brane reheating temperature does not depend on the initial value
of the inflaton field and is solely determined by the elementary particle theory of the early universe.


\subsection{Phase transition in brane inflation}
Phase transition in braneworld scenario is weakly
first order in nature \cite{Narlikar}. So it is convenient to write the
brane reheating temperature in terms of the critical parameters. To serve this purpose
the critical density and the critical temperature or transition temperature can be written as :
\be\label{roz}\rho(t_{c})=2\lambda=\frac{3}{16\pi^{2}}\frac{M^{6}_{5}}{M^{2}},~~~T_{c}=\sqrt{\frac{3}{\pi}\sqrt{\frac{5}{\pi
N^{*}}}\frac{M^{3}_{5}}{M_{PL}}}\ee
which makes a bridge between the phenomenology and observation. In the high energy limit 5D Planck mass ($M_{5}$)
can be expressed in terms of our model parameters as
 $M_{5}=\sqrt[6]{\frac{6400\pi^{4}\Delta^{2}_{s}(K_{4}+4D_{4})^{2}}{\alpha^{4}}}\phi_{\star}$.
Here $\Delta^{2}_{s}$ represents the amplitude of the scalar perturbation defined as $\Delta^{2}_{s} \simeq
\frac{512\pi}{75M^{6}_{PL}}\left[\frac{V^{3}}
{(V^{'})^{2}}\left[1+\frac{V}{2\lambda}\right]^{3}\right]_{\star}$. Most importantly here the subscript $\star$
represents here the epoch of horizon crossing ($k=aH$) and $\alpha$ represents a dimensionless model parameter defined as $\alpha=\frac{\Delta^{4}}{\lambda}$.

The major thermodynamic quantities -- critical density ($\rho_{c}$), critical pressure ($P_{c}$), critical entropy ($S_{c}$)
 -- and the Hubble parameter at the critical temperature ($H_{c}$) related to the phase
transition designated by a four tuple critical characteristic set $U(c\gamma)$ by the following fashion for our model:
\be\begin{array}{ll}\label{comp} \displaystyle U(c\gamma):\left[\left\{\rho_{c},P_{c},S_{c},H_{c}\right\}\right.
\\ \left.\displaystyle\equiv\left[\phi^{4}_{\star}A(\phi_{\star})\left\{1200,400,\frac{1600}{T_{c}},\frac{20}{\sqrt{A(\phi_{\star})}M\phi^{2}_{\star}}\right\}\right]\forall \gamma\in J\right]\end{array}\ee
where we have defined a dimensionless characteristic quantity $A(\phi_{\star})=\frac{\pi^{2}(K_{4}+4D_{4})^{2}\Delta^{2}_{s}\phi^{2}_{\star}}{\alpha^{4}M^{2}}$ at the horizon crossing in this context.
 The above mentioned physical
 quantities are function of the critical or transition temperature which is defined as
\be\begin{array}{ll}\label{ubi}\displaystyle T_{c}:=\left[T_{c\gamma}=\sqrt[4]{\left\{C_{\gamma}
\frac{A(\phi_{\star})\phi^{4}_{\star}}{\pi^{2}N^{*}_{\gamma}}\right\}}
\right. \\ \displaystyle\left.~~ with ~~ C_{\gamma}=\left(36000,\frac{288000}{7},19200\right)\forall \gamma\in J\right]\end{array}\ee
 with gauge group $J:=SU(2)_{L}\otimes U(1)_{Y}$ and the species index $\gamma=1(B\Rightarrow Boson),2(F\Rightarrow Fermion),
3(M\Rightarrow Mixture)$.

\begin{figure}[htb]
{\centerline{\includegraphics[width=8cm, height=6cm] {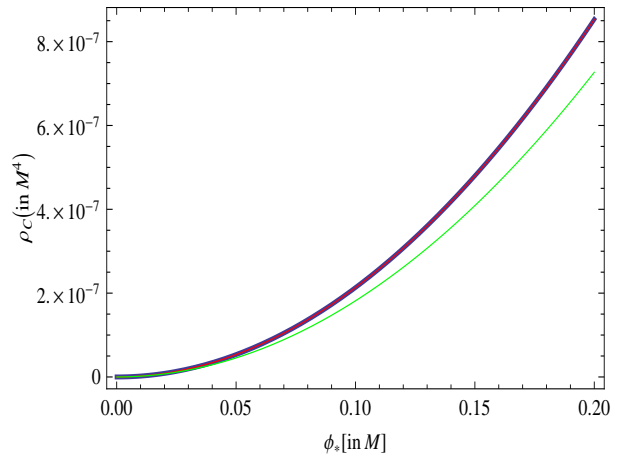}}}
\caption{Here we have plotted the variation of the  critical density with respect to the effective inflaton field $\phi_{\star}$
after horizon crossing in the domain $-0.70<D_{4}<-0.60$,
which explicitly shows
 the direct connection between the phenomenological and observational sector through the 5D Planck mass $M_{5}$ given in
eqn(\ref{roz}). In addition it
 confirms the existence of phase transition in braneworld scenario in high energy limit.
Here we have used $M=2.43\times 10^{18}GeV$. Most importantly the energy density is smoothly varying function of inflaton field
in the critical domain.} \label{figVr133}
\end{figure}

\subsection{Brane reheating temperature}

 In this context the reheating temperature can be written \cite{felipe} as a one to one mapping ($\O$) in parameter space as \be\begin{array}{ll}\label{reh1}
\displaystyle\O :\left[\left\{T^{br}:=\frac{T_{c}}{\sqrt[4]{2}}\sqrt[4]{\left[\sqrt{1+\frac{5}{\pi^{3}N^{*}}
\left(\frac{\Gamma_{total}M_{PL}}{T^{2}_{c}}\right)^{2}}-1\right]}\right.\right.
\\ \left.\left.~~~~\displaystyle\Rightarrow T^{brh}=\sqrt[4]{\left\{\sqrt{\frac{10}{N^{*}}}\frac{2\sqrt{2}M
\Gamma_{total} T^{2}_{c}}{3\pi}\right\}}\right\}\in \c{C}\right]
\end{array}\ee
where $\c{C}$ represents collection of all gauge group which supports particle theory. But in this context we are confining ourselves into
the Standard Model regime. So to construct a fruitful model of reheating in the context of Standard Model gauge group, we rewrite all general principal components in terms of physical degrees of freedom in a compact fashion. We consider a one to one high energy mapping $Q[\gamma]$ in a physical space such that
\be\begin{array}{ll}\label{f1}\displaystyle Q[\gamma]:\left\{\left[T^{br}_{\gamma}=\frac{T_{c\gamma}}{\sqrt[4]{2}}\sqrt[4]{\left[\sqrt{1+\frac{Z_{\gamma}}{\pi^{3}N^{*}_{\gamma}}\left(\frac{\Gamma_{total}M_{PL}}{T^{2}_{c\gamma}}\right)^{2}}-1\right]}\right.\right.
\\ \left.\left.~~~~\displaystyle\Longrightarrow
T^{brh}_{\gamma}=\sqrt[4]{\left\{\frac{W_{\gamma}(K_{4}+4D_{4})\Delta_{s}\phi^{3}_{\star}\Gamma_{total}}{\pi
N^{*}_{\gamma}\alpha^{2}}\right\}}\right]\forall \gamma \in J\right\}\end{array}\ee it maps the actual brane reheating temperature ($T^{br}_{\gamma}$) to its high energy value ($T^{brh}_{\gamma}$) in the Standard Model gauge group $J:=SU(2)_{L}\otimes U(1)_{Y}$
with $Z_{\gamma}=\left(5,\frac{40}{7},\frac{8}{3}\right)$,$W_{\gamma}=\left(600,\frac{4800}{7},320\right)$ and $\gamma=1(B),2(F),3(M)$. Here
$\bigcup_{\gamma}U[c\gamma]\bigoplus Q[\gamma]\subseteq \O$ for which $J \in \c{C}$.
Most importantly the superscript `br' and `brh' stands for parameters before and after high energy mapping respectively.
 Here it should be mentioned that the
brane reheating temperature incorporates all the effects of heavy Majorana neutrinos as well as the other fermions and bosons
through the total decay width $\Gamma_{total}$.

The reheating temperature for different species can readily be calculated from our model. For a typical value of $C_{4}
\simeq D_{4}=-0.7$ (consistent with \cite{sayan}), we have: for boson $T^{brh}_{B}\simeq7.6\times 10^{10}GeV$, for fermion 
$T^{brh}_{F}\simeq7.8\times 10^{10}GeV$ and
 for mixture of species $T^{brh}_{M}\simeq6.5\times 10^{10}GeV$. This is
significantly different from GR value $T^{reh}\leq10^{6}-10^{7}GeV$ and is a characteristic feature of brane inflation.  


\section{Gravitino phenomenology  on the brane for $\bf SU(3)_{C}\bigotimes SU(2)_{L}\bigotimes U(1)_{Y}$}

Let us now move on to studying how the self interacting term of our model is directly related to the
leptogenesis through the production of thermal gravitinos which is a special ingredient for the heavy  Majorana neutrinos
in the leptogenesis. Let us start with a physical situation where the inflaton field starts
oscillating when the inflationary epoch ends at a cosmic time
$t=t_{osc}\simeq t_{f}$. Throughout the analysis we have assumed that the
universe is reheated through the perturbative decay of the
inflaton field for which the reheating phenomenology in brane is described
by the Boltzmann equation \cite{turner}
\be\label{rfg} \dot{\rho_{r}}+4H\rho_{r}=\Gamma_{\phi}\rho_{\phi},\ee
where in braneworld
\be\begin{array}{ll}\label{rgg1} \displaystyle H^{2}=\frac{8\pi}{3M^{2}_{PL}}\left(\rho_{r}+\rho_{\phi}\right)\left[1+\frac{(\rho_{r}+\rho_{\phi})}{2\lambda}\right]
\\ \displaystyle =H^{2}_{osc}\left(\frac{a_{osc}}{a}\right)^{4}\left[1+\frac{\alpha}{2}\left(\frac{a_{osc}}{a}\right)^{4}\right].\end{array}\ee
Here $\rho_{r}$ and $\rho_{\phi}$ represent the energy density of radiation and inflaton respectively and
$\Gamma_{\phi}$ is the rate of dissipation of the inflaton field energy
density. At $t=t_{osc}$ epoch the Hubble parameter is designated by \cite{turner}\be\label{ddg}
H_{osc}=\sqrt{\frac{8\pi}{3}}\frac{\Delta^{2}}{M_{PL}}=\frac{\Delta^{2}}{\sqrt{3}M}.\ee
 Assuming $\Gamma_{\phi}\gg H$ from we get \be\label{gh1}
\rho_{\phi}=\Delta^{4}\left(\frac{a_{osc}}{a}\right)^{4}\exp\left[-\Gamma_{\phi}(t-t_{osc})\right].\ee
It is worthwhile to mention here that the inflaton field $\phi$ follows
 an equation of state similar to radiation rather than matter i.e. $\omega_{\phi}=\frac{P_{\phi}}{\rho_{\phi}}$,
where $P_{\phi}=\rho_{\phi}-\Delta^{4}\left[1+\left(D_{4}+K_{4}\ln\left(\frac{\phi}{M}\right)\right)
\left(\frac{\phi}{M}\right)^{4}\right]$.
Now solving Friedmann equation the dynamical character of the scale factor can be expressed as \be\label{dt1}
a(t)=a_{osc}\sqrt[4]{\left[\left[\sqrt{1+\frac{\alpha}{2}}+2H_{osc}(t-t_{osc})\right]^{2}-\frac{\alpha}{2}\right]},\ee
where we use a specific notation $a(t_{osc})=a_{osc}$.

  Plugging eqn(\ref{dt1}) and eqn(\ref{gh1}) in eqn(\ref{rfg})
we get \be\begin{array}{ll}\label{opp}
\displaystyle\dot{\rho_{r}}+\frac{2H_{osc}}{\left[\left[\sqrt{1+\frac{\alpha}{2}}+2H_{osc}(t-t_{osc})\right]^{2}-\frac{\alpha}{2}\right]}\rho_{r}
\\ \displaystyle=\frac{\Gamma_{\phi}\Delta^{4}\exp\left[-\Gamma_{\phi}(t-t_{osc})\right]
}{\left[\left[\sqrt{1+\frac{\alpha}{2}}+2H_{osc}(t-t_{osc})\right]^{2}-\frac{\alpha}{2}\right]},\end{array}\ee
 As a whole phenomenological construction of gravitino abundance is governed by the above equation. But eqn(\ref{opp})
is not exactly analytically solvable. So we are confining our attention to the high energy limit where
the Friedmann equation (\ref{rgg1}) can be approximated as
\be\label{ty} H^{2}=\frac{8\pi}{6\lambda
       M^{2}_{PL}}\left(\rho_{r}+\rho_{\phi}\right)^{2}=\frac{\alpha}{2}H^{2}_{osc}\left(\frac{a_{osc}}{a}\right)^{8},\ee
     whose solution is given by \be\label{olpiuy}a(t)=a_{osc}\sqrt[4]{\left[1+2\sqrt{2\alpha}H_{osc}(t-t_{osc})\right]}.\ee

     Now using an physically viable assumption $t\leq \Gamma^{-1}_{\phi}$ the exact solution of the eqn(\ref{opp}) in the high energy limit can be written as \bea\label{sd1}
        \rho_{r} &\simeq& \frac{3M^{2}H^{2}_{osc}\Gamma_{\phi}(t-t_{osc})}{\left[1+2\sqrt{2\alpha}H_{osc}(t-t_{osc})\right]}
            \nonumber   \\  &=&\frac{3M^{2}H_{osc}\Gamma_{\phi}}{2\sqrt{2\alpha}}\left(\frac{a_{osc}}{a}\right)^{4}\left[\left(\frac{a}{a_{osc}}\right)^{4}-1\right].\eea

Our intention is to find out the extremum temperature during
reheating epoch which is one of the prime components for the determination of gravitino abundance. In the braneworld scenario this extremum temperature
is given by \be\label{cv1}
 T^{bh}_{ex}=\sqrt[4]{\left[\frac{13\sqrt{3}\Delta^{2}M\Gamma_{\phi}}{N^{*}\pi^{2}}\sqrt{\frac{1}{2\alpha}}\right]}
=\sqrt[4]{\left\{\frac{45\Gamma_{\phi}M^{3}_{5}}{8N^{*}\pi^{3}}\right\}}\ee
and it is less than the reheating temperature in brane ($T^{brh}$). This phenomenon is different from standard GR results \cite{raga} where we see that the reheating temperature shoots up to a maximum value and it gives the upper
bound of the reheating temperature. But in the present context of brane inflation this situation is completely different i.e. at first temperature falls down to a minimum which fixes the lower bound of the reheating temperature and rises to a maximum at the end of reheating
 epoch. Using eqn(\ref{opp}), eqn(\ref{cv1}) and the thermodynamic background of energy density
 of radiation we can express the scale factor in terms of temperature as \be\label{xc1}
 a(T)=
\left\{
	\begin{array}{ll}
                    \frac{a_{osc}}{\sqrt[4]{\left[1-32\left(\frac{T}{T^{bh}_{ex}}\right)^{4}\right]}}& \mbox{if } t=t_{osc}\simeq t_{f} \\
           \frac{a_{osc}}{\sqrt[4]{\left[32\left(\frac{T}{T^{bh}_{ex}}\right)^{4}-1\right]}}& \mbox{if } t_{osc}(\simeq t_{f})<t\leq t_{reh}.
          \end{array}
\right.
\ee
 It is worth mentioning that if we break the time scale into two parts $t_{osc}<t<t_{ex}$ and $t_{ex}<t<t_{reh}$,
as done in GR the scale
factor and hence the remaining results have same expressions in these two different zones. This is in sharp contrast with standard GR
results except at $t=t_{f}$, where they have different values in the two different regimes.

\begin{figure}[htb]
{\includegraphics[width=8cm, height=6cm] {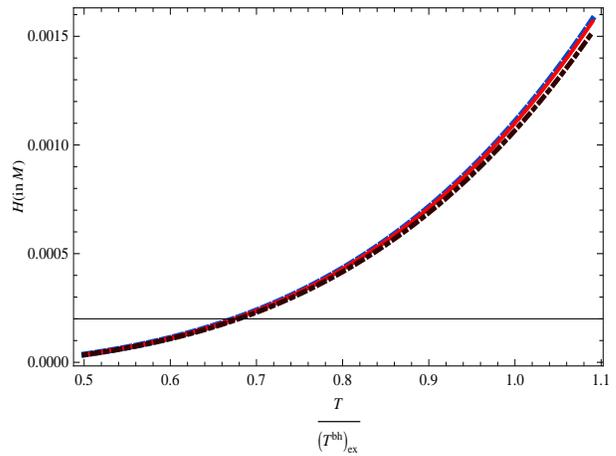}}
\caption{In the above figure we have plotted the variation of the Hubble parameter
  with respect to dimensionless parameter $\frac{T}{T^{bh}_{ex}}$ in the domain $-0.70<D_{4}<-0.60 $,
which shows the smooth behavior of Hubble parameter except $x\leq 0$ i.e. at $\frac{T}{T^{bh}_{ex}}\leq\frac{1}{\sqrt[4]{32}}$.
Most importantly here the equality corresponds to the end of reheating epoch and the beginning of radiation dominated era which is the
direct outcome of the
first expression at $t\simeq t_{f}$ for the scale factor ($a(T)$) stated in eqn(\ref{xc1}). The rest of the the part
follows the second expression given in eqn(\ref{xc1}) in the interval $t_{f}<t<t_{reh}$  plotted in the above figure.
Additionally the vertical scale corresponds to $M=2.43\times 10^{18}GeV$.} \label{figVr400}
\end{figure}


 Let us now use this phenomenological background to derive the expression of the gravitino production during two thermal epochs -  reheating and radiation dominated era. It is well known that gravitinos are produced by the scattering of the inflaton decay products \cite{ferran}. The master equation of gravitino phenomenology as obtained from `Boltzmann equation.' is given by \cite{prad,stef}
  \be\label{bn1}
\frac{dn_{\tilde{G}}}{dt}+3Hn_{\tilde{G}}=\langle\Sigma_{total}|v|\rangle n^{2}-\frac{m_{\frac{3}{2}}n_{\tilde{G}}}{\langle E_{\frac{3}{2}}\rangle\tau_{\frac{3}{2}}},\ee where $n=\frac{\zeta(3)T^{3}}{\pi^{2}}$ is the number density of scatterers(bosons in thermal bath) with
$\zeta(3)$=1.20206.... Here $\Sigma_{total}$ is the total scattering cross section for thermal gravitino production, $v$ is the relative velocity of the incoming particles with $\langle v\rangle=1$ where $\langle...\rangle$ represents the thermal average. The factor $\frac{m_{\frac{3}{2}}}{\langle E_{\frac{3}{2}}\rangle}$ represents the averaged Lorentz factor which comes from the decay of gravitinos can be neglected due to weak interaction.
 For the gauge group $E:=SU(3)_{C}\bigotimes SU(2)_{L}\bigotimes U(1)_{Y}$ the thermal gravitino production rate is given by, \be\begin{array}{ll}\label{mn1}
\displaystyle\langle\Sigma_{total}|v|\rangle=\frac{\tilde{\alpha}}{M^{2}}\\~~~~~~~~~~~~~\displaystyle=\frac{3\pi}{16\zeta(3)M^{2}}\sum^{3}_{i=1}\left[1+\frac{M^{2}_{i}}{3m^{2}_{\tilde{G}}}\right]
C_{i}g^{2}_{i}\ln\left(\frac{K_{i}}{g_{i}}\right),\end{array}\ee where $i=1,2,3$ stands for the three gauge groups $U(1)_{Y}$,$SU(2)_{L}$ and $SU(3)_{C}$ respectively. Here $M_{i}$ represent gaugino mass parameters and $g_{i}(T)$ represents gaugino coupling constant at finite temperature (from MSSM RGE)\be\label{hh1}
g_{i}(T)\simeq\frac{1}{\sqrt{\frac{1}{g^{2}_{i}(M_{Z})}-\frac{b_{i}}{8\pi^{2}}\ln\left(\frac{T}{M_{Z}}\right)}}\ee
with $b_{1}=11,b_{2}=1,b_{3}=-3$. Here $C_{i}$ and $K_{i}$ represents the constant associated with the gauge groups with $C_{1}=11,C_{2}=27,C_{3}=72$ and $K_{1}=1.266,K_{2}=1.312,K_{3}=1.271$.

For convenience let us recast eqn(\ref{bn1}) as
 \be\label{uiu}\dot{T}\frac{dn_{\tilde{G}}}{dT}+3Hn_{\tilde{G}}=\langle\Sigma_{total}|v|\rangle
n^{2},\ee where a boundary condition $T=T^{bh}_{ex}$ ,$\dot{T}=0$ is introduced.
In terms of a dimensionless variable \be\label{az1} x=32\left(\frac{T}{T^{bh}_{ex}}\right)^{4}-1\ee
 eqn(\ref{uiu}) can be expressed as \be\label{iop}\frac{dn_{\tilde{G}}}{dx}+\frac{d_{1}}{x}n_{\tilde{G}}=-\frac{d_{3}(x+1)^{\frac{3}{2}}}{x^{2}}\ee
where $ d_{1}=-\frac{3}{4},~~~d_{3}=\frac{(T^{bh}_{ex})^{6}}{32}
\frac{\tilde{\alpha}}{M^{2}}\left(\frac{\zeta(3)}{\pi^{2}}\right)^{2}\frac{\sqrt{\lambda}}{4\sqrt{3}H^{2}_{osc}M}$.
The exact solution of the eqn(\ref{iop}) is given by
\be
\begin{array}{ll}
\label{d1}n_{\tilde{G}}(x)=\frac{2d_{3}}{x^{d_{1}}}\sqrt{x+1}\left(-2\,_2F_1\left[\frac{1}{2};1-d_{1};\frac{3}{2};x+1\right]\right.
 \\ \left.+_2F_1\left[\frac{1}{2};2-d_{1};\frac{3}{2};x+1\right] +_2F_1\left[\frac{1}{2};-d_{1};\frac{3}{2};x+1\right]\right),
\end{array}
\ee
Using the properties of Gaussian hypergeometric function for $x>>1$ eqn(\ref{d1}) reduces to the following simpler form:
\be\begin{array}{ll}
    \label{zac}n_{\tilde{G}}(x)\simeq2d_{3}x^{\frac{1}{4}}\sqrt{1+x}\frac{\Gamma(\frac{3}{2})\Gamma(\frac{1}{2})}{\Gamma(1)}
   \left\{\frac{1}{\Gamma(\frac{3}{4})}+\frac{1}{\Gamma(-\frac{5}{4})}-\frac{2}{\Gamma(-\frac{1}{4})}\right\}\end{array}
\ee
Using the boundary condition $T=T^{bh}_{ex}$ in eqn(\ref{zac}) the numerical value of gravitino abundance turns out to be
$n_{\tilde{G}}(x_{ex})=62.023d_{3}$.

\begin{figure}[htb]
{\includegraphics[width=8cm, height=6cm] {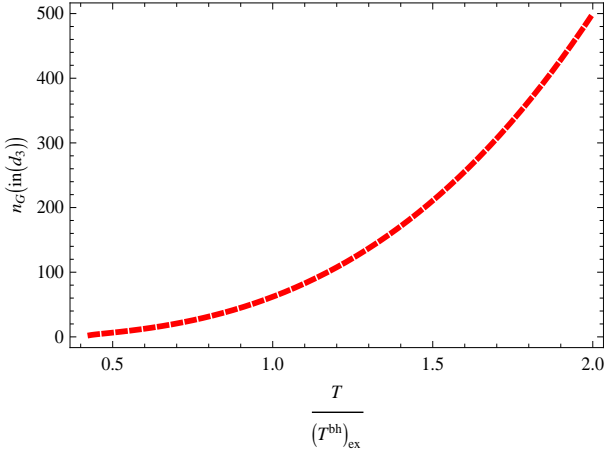}}
\caption{In the above diagram we have plotted variation of gravitino number density in a physical volume  vs scaled temperature in braneworlds.
Here we have used the fundamental scale $d_{3}=4.596\times10^{-44}\tilde{\alpha}M^{3}$
, where $\tilde{\alpha}$ is a dimensionless number depends on the species of the MSSM gauge group $E$. For an
example $n=4$ level flat direction content  ${\bf QQQL, QuQd, QuLe, uude }$ of MSSM gives $\tilde{\alpha}\simeq 15.694$ in the absence
of top Yukawa coupling. Most importantly
4D effective Planck mass $M=2.43\times
10^{18}GeV$. From the plot it is obvious that the gravitino number density is monotonically increasing function of the dimensionless
variable $\frac{T}{T^{bh}_{ex}}$ except at $x\leq 0 $ which implies $\frac{T}{T^{bh}_{ex}}\leq\frac{1}{\sqrt[4]{32}}$. } \label{figVr900}
\end{figure}

Let us now find out the exact analytical expression for the gravitino abundance at reheating temperature $T^{brh}$ in the high energy limit. To serve this purpose substituting $T=T^{brh}$ in eqn(\ref{d1}) we get
\be\begin{array}{ll}\label{bhu}
  \displaystyle n_{\tilde{G}}(T^{brh})=8\sqrt{2}d_{3} \left(32\left(\frac{T^{brh}}{T^{bh}_{ex}}\right)^{4}-1\right)^{\frac{3}{4}}
\\ \displaystyle~~~~~~~~~~~~~~~~~\times \left(\frac{T^{brh}}{T^{bh}_{ex}}\right)^{2}G\left(\frac{T^{brh}}{T^{bh}_{ex}}\right)
 \end{array}\ee
where \be\begin{array}{ll}\label{sctt}  G\left(\frac{T^{brh}}{T^{bh}_{ex}}\right)=\left(-2\,_2F_1\left[\frac{1}{2};\frac{7}{4};\frac{3}{2};32\left(\frac{T^{brh}}{T^{bh}_{ex}}\right)^{2}\right]
   \right.\\ \left.+_2F_1\left[\frac{1}{2};\frac{11}{4};\frac{3}{2};32\left(\frac{T^{brh}}{T^{bh}_{ex}}\right)^{2}\right]
   +_2F_1\left[\frac{1}{2};\frac{3}{4};\frac{3}{2};32\left(\frac{T^{brh}}{T^{bh}_{ex}}\right)^{2}\right]\right)
\end{array}\ee
along with an extra constraint $G\left(\frac{T^{brh}}{T^{bh}_{ex}}>>\frac{1}{\sqrt[4]{32}}\right)
=\frac{\pi}{2}\left(32\left(\frac{T^{brh}}{T^{bh}_{ex}}\right)^{4}-1\right)^{-\frac{1}{2}}
 \left\{\frac{1}{\Gamma(\frac{3}{4})}+\frac{1}{\Gamma(-\frac{5}{4})}-\frac{2}{\Gamma(-\frac{1}{4})}\right\} $.
 It is convenient to express the abundance of any species `$\sigma$'  as\cite{raga} $ Y_{b}=\frac{n_{b}}{s}$ where $n_{b}$
 is the number density of the species `$b$' in a physical volume and `s' is the entropy density given by $s=\frac{2\pi^{2}}{45}N^{*}T^{3}$.
 Here the master equation. for gravitino can be expressed as \be\label{zx1}
 \dot{T}\frac{dY^{br}_{\tilde{G}}}{dT}=\langle\Sigma_{total}|v|\rangle Y^{br}_{\tilde{G}}n\ee

Using eqn(\ref{dt1}) and eqn(\ref{xc1}) the time-temperature relation can be found as:
\be\label{bnmj}T=\frac{T^{br}}{\sqrt[4]{\left[\left[\sqrt{1+\frac{\alpha}{2}}+2H_{reh}
(t-t_{reh})\right]^{2}-\frac{\alpha}{2}\right]}}.\ee

Eliminating $\dot{T}$ we get the solution of the master equation(\ref{zx1}) in the radiation dominated era as \be\label{ft}
 Y^{br}_{\tilde{G}}(T_{f})=Y^{br}_{\tilde{G}}(T^{br})+Y^{br-rad}_{\tilde{G}}(T_{f})\ee
 where\be\begin{array}{ll}\label{dct2}\displaystyle Y^{br-rad}_{\tilde{G}}(T_{f})=\sqrt{\frac{90}{\pi^{2}N^{*}}}\left(\frac{45\sqrt{2}}{2\pi^{2}N^{*}\sqrt{\alpha}}\right)\left(\frac{\tilde{\alpha}}{M}\right)\left(\frac{\zeta(3)}{\pi^{2}}\right)^{2}
 \\ ~\displaystyle\times\frac{T^{br}}{T_{f}\sqrt{1+\frac{\pi^{2}}{60\lambda}N^{*}(T^{br})^{4}}}
\left(T^{br}\,_2F_1\left[\frac{1}{4};\frac{1}{2};\frac{5}{4}; -\frac{2(T^{br})^{4}}{\alpha T^{4}_{f}}\right]\right.
\\ \left.~~~~~~~~~~~~~~~~~~~~~~~~~~~~~~~~~~\displaystyle-T_{f}\,_2F_1\left[\frac{1}{4};\frac{1}{2};\frac{5}{4};-\frac{2}{\alpha}\right]\right)\end{array}\ee
 But in eqn(\ref{ft}) the first term on the right-hand side is not
 exactly computable. As mentioned earlier to find out exact expression we have used here the high energy mapping.

In the radiation dominated era the dynamical behavior of temperature can be mapped as
 \be\begin{array}{l}\label{hj2}\Gamma:\displaystyle\left[\left\{T=
\left(\frac{T^{br}}{\sqrt[4]{\left[\left[\sqrt{1+\frac{\alpha}{2}}+2H_{reh}
(t-t_{reh})\right]^{2}-\frac{\alpha}{2}\right]}}\right.\right.\right.
\\ \left.\left.\left.~~~~~~\displaystyle\Longrightarrow \frac{T^{brh}}{
\left[1+2\sqrt{2\alpha}H_{reh}(t-t_{reh})\right]^{\frac{1}{4}}}\right)\right\}\in E\right]\end{array}\ee

Using this map we finally have
\be\begin{array}{ll}\label{fgh4}
     Y^{brh}_{\tilde{G}}(T_{f})=\left(\frac{\tilde{\alpha}}{M}\right)
     \left(\frac{\zeta(3)}{\pi^{2}}\right)^{2}\left(\frac{45\sqrt{3\lambda}}{2\pi^{3}\Delta^{2}N^{*}}\right) \left[\left(\frac{60\sqrt{\lambda}}{\pi N^{*}T_{f}}\right)
     \left(1-\frac{T_{f}}{T^{brh}}\right)\right.\\ \left.+\left(\frac{(T^{bh}_{ex})^{4}}{16\Delta^{2}T^{brh}}\right)
\left(32\left(\frac{T^{brh}}{T^{bh}_{ex}}\right)^{4}-1\right)^{\frac{3}{4}}G\left(\frac{T^{brh}}{T^{bh}_{ex}}\right)\right].\end{array}\ee

where
 \be\begin{array}{l}\label{yh2}
     Y^{b-rad}_{\tilde{G}}(T_{f})=\left(\frac{6\tilde{\alpha}}{M}\right)\left(\frac{\zeta(3)}{\pi^{2}}\right)^{2}
     \sqrt{\frac{3\lambda}{\alpha}}\left(\frac{15}{\pi^{2}N^{*}}\right)^{2}\left(\frac{1}{T_{f}}-\frac{1}{T^{brh}}\right),\end{array}\ee
     and \be\begin{array}{l}\label{rd6} Y^{brh}_{\tilde{G}}\simeq Y^{b}_{\tilde{G}}=\frac{n_{\tilde{G}}}{s}=\left(\frac{360\sqrt{2}d_{3}}{2\pi^{2}N^{*}(T^{brh})^{3}}\right)\left(32\left(\frac{T^{brh}}{T^{bh}_{ex}}\right)^{4}-1\right)^{\frac{3}{4}}
         \\ ~~~~~~~~~~~~~~~~~~~~~~~~~~~~~~~\times\left(\frac{T^{brh}}{T^{bh}_{ex}}\right)^{2}G\left(\frac{T^{brh}}{T^{bh}_{ex}}\right)\end{array}.\ee
The gravitino dark matter abundance and
the baryon asymmetry is connected through $Y^{brh}_{\tilde{G}}\simeq\frac{\Theta_{CP}{\cal D}}{N^{*}}$
 where ${\cal D}(\leq 1)$ is the
dilution factor and the leading contribution is given
by the interference between the tree level and the one-loop level decay amplitudes. Here the
CP-violating parameter is described as \cite{okada}
$\Theta_{CP}=
\frac{\Gamma(\psi\rightarrow\bar{l}_{L}{\cal H})
-\Gamma(\psi\rightarrow l_{L}{\cal H^{\star}})}{\Gamma(\psi\rightarrow\bar{l}_{L}{\cal H})
+\Gamma(\psi\rightarrow l_{L}{\cal H^{\star}})}=\frac{3{\cal M}m_{\nu}}{16\pi v^2}\sin\delta_{CP}$,
where $m_{\nu}$ is the heaviest light neutrino mass, $v = 174 GeV$ is the VEV of Higgs and $\delta_{CP}$ is an effective CP phase
 which parameterize each entries of the CKM matrix. Particularly $\delta_{CP}$ acts as a probe of
 flavor structure in supergravity theories.
The complete wash out situation corresponds to ${\cal D}=1$.

\begin{figure}[htb]
{\includegraphics[width=8.5cm, height=6cm] {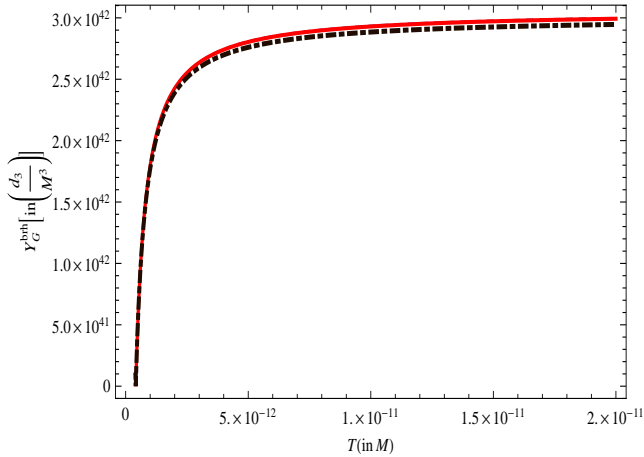}}
\caption{
Here we have plotted the variation of total gravitino abundance vs temperature in the domain $-0.70<D_{4}<-0.60$,
which clearly shows that  gravitino abundance  at zero temperature shoots up initially to maximum and then becomes
 constant with respect to temperature during radiation dominated era in braneworld scenario.
As mentioned earlier we have used the fundamental scale $\frac{d_{3}}{M^{3}}=4.596\times10^{-44}\tilde{\alpha}$
, where $\tilde{\alpha}$ is a dimensionless number which depends on the species of the MSSM and $M=2.43\times
10^{18}GeV$. } \label{figVr600}
\end{figure}

 Through out all the numerical estimation we have taken decay width
 $\Gamma_{\phi}\simeq2.9\times 10^{-3}GeV$,  mass of the
inflaton $m_{\phi}\simeq10^{13}GeV$, final temperature and time at the end of reheating $T_{f}\simeq10^{6}GeV$ and
 $t_{f}\simeq1.4\times 10^{31}GeV$ respectively.  For a typical value of $C_{4}\simeq D_{4}=-0.7$
 extremum (minimum) temperature during reheating can be estimated as
 $ T^{bh}_{ex}\simeq7.0\times10^{10}GeV$. This clearly
 shows deviation from standard GR phenomenology \cite{raga}
 where the extremum (maximum) temperature during reheating $T_{max}\simeq 1.3\times 10^{12}GeV$. Similarly 
for $C_{4}\simeq D_{4}=-0.7$ the critical temperature for different particle species and gravitino abundance at different temperatures 
obtained from our model are: for boson $T_{cB}\simeq3.2\times 10^{14}GeV$, for fermion $T_{cF}\simeq3.3\times 10^{14}GeV$,
 for mixture of species $T_{cM}\simeq2.8\times 10^{14}GeV$, at reheating temperature $Y^{b}_{\tilde{G}}(T^{brh})\simeq8.1\times 10^{-34}
GeV^{-3}d_{3}$ and at the end of reheating $Y^{b-rad}_{\tilde{G}}(T_{f})\simeq2.1\times 10^{-13}GeV^{-3}d_{3}$.
 We have calculated all the abundances in the fundamental unit of
$d_{3}$ i.e. $d_{3}=6.594\tilde{\alpha}\times10^{11}GeV^{3}$, where $\tilde{\alpha}$ is a dimensionless
characteristic constant originated through the thermal gravitino production rate in the context of MSSM. Most significantly for
 different flat direction contents the phenomenological parameter is different and can be calculated
from MSSM RGE flow at the one-loop level for that flat direction. To obtain a conservative
estimate of gravitino abundance we have taken here gaugino masses $M_{i}\rightarrow 0$ for all gauge subgroups within MSSM.
 For example the fourth level flat directions
${\bf QQQL, QuQd, QuLe, uude }$ give $\tilde{\alpha}=15.694$ for a specific choice of the
$U_{Y}(1)$, $SU_{L}(2)$ and $SU_{C}(3)$ gauge couplings $g_{1}=0.56$, $g_{2}=0.72$ and $g_{3}=0.85$ respectively
 obtained from the {\it universal mSUGRA boundary condition} and consistent with electroweak extrapolation of the solution of MSSM RGE
flow from the energy scale of brane inflation $\Delta=0.2\times10^{16}GeV$ for our model.
The linear dependence on $T^{brh}$ makes simple to revise
the constraints on $T^{brh}$ based on the lower limit on the
gravitino abundance - the lower bound on  $T^{brh}$ is increased by a factor of 1.074. Since $T^{bh}_{ex} \propto T^{brh}$,
 $T^{bh}_{ex}$ is
not affected much. Therefore models of leptogenesis that
invoke a small $T^{bh}_{ex}$ to create heavy Majorana neutrinos
are not significantly affected.
 Within $55<N<70$  and  $T^{brh}\simeq6.5\times 10^{10}GeV$  the entropy density
changes. As a consequence the total gravitino abundance changes according to fig(\ref{figVr600}).
It is easily seen that $P=\frac{\rho}{3}$, $S=\frac{\rho+P}{T}$ consistency relations are valid in this context.
 It is worthwhile to mention here that in brane pressure and entropy density of the universe falls down to a minimum due to the minimum temperature during
reheating epoch. However during radiation dominated era total entropy density is almost constant for both the cases. This clearly shows
the deviation from standard GR phenomenology. Throughout the analysis we have not included the effect of $\exp[-\Gamma_{\phi} (t - t_{osc})]$
in the energy density of inflaton $\rho_{\phi}$. One might be concerned that this
will lead to inaccuracies close to $t_{brh}$ when most of
the gravitinos are produced. However if one writes
$\rho_{\phi}\simeq a^{-4} \exp(-\Gamma_{\phi} t)\simeq t^{-2}\exp(-\Gamma_{\phi} t)$ for $t >> t^{bh}_{ex}$
then $\dot{\rho_{\phi}}/\rho_{\phi} = -2/t -\Gamma_{\phi}$. Therefore even till close to
 ̇$t_{brh}=\Gamma^{-1}_{\phi}\rho_{\phi}$ decreases primarily due to the expansion
of the universe. Furthermore, near $t_{brh}$ it increases as $T^{-1/2}\simeq \sqrt{a}\simeq t^{\frac{1}{8}}$ in brane
 which is again different from GR phenomenology where $T^{-1/2}\simeq \sqrt{a}\simeq t^{\frac{1}{4}}$.
The thermal leptogenesis in the braneworld can take place if the lightest heavy
neutrino mass lying in the range
$T^{brh} < {\cal M} < T_{c}$. This confirms that the
upper bound of 5D Planck mass $M_{5}< 10^{16}GeV$ (for our model $M_{5}\simeq7.8\times 10^{15}GeV$ for $C_{4}\simeq D_{4}=-0.7$),
 which coincides with the
leptogenesis bound implied by the observed baryon asymmetry. It is important to mention here that in the standard cosmology,
 the thermal leptogenesis in supergravity models is hard to be successful,
 since the reheating temperature after inflation is severely
constrained to be $T^{reh}\leq10^{6}-10^{7}GeV$ due to the gravitino problem. However, as pointed
out in \cite{felipe}, the constraint on the reheating temperature is replaced by
the transition temperature in the brane world cosmology. As a result the gravitino problem can be solved even if the
reheating temperature is much higher. In fact, such inflation
models are possible but limited and our model is also in that category. Here we are using a preferable
 value of the heaviest light neutrino mass from
atmospheric neutrino oscillation data
$m_{\nu} \simeq 0.05 eV$ and for sufficient
 baryon asymmetry the lightest neutrino mass ${\cal M}\simeq 10^{10}GeV$. For complete washout situation (${\cal D}=1$) in our model the effective CP phase lying within the window
 $2.704\times10^{-9} <\delta_{CP}< 2.784\times10^{-9}$, where $\delta_{CP}$ is measured in degree.
 Most significantly it indicates that the amount
of CP violation in braneworld scenario is very small and identified with the {\it soft CP phase}. Consequently it has negligibly small
contribution to ${\cal K}$ and ${\cal B}$ physics phenomenology.


\section{Summary and outlook}

In the present article we have studied reheating in brane
cosmology on the background of supergravity. We have  exhibited
 the process of construction of a fruitful theory of reheating for an effective
4D inflationary potential in $N=1, D=4$ supergravity in the brane derived from  $N=2, D=5$ supergravity in
the bulk \cite{sayan}. We have employed this
potential in reheating model building by analyzing the reheating temperature in the context of brane inflation, followed
by analytical and numerical estimation of different phenomenological parameters.
 It is worthwhile to mention here that we get a lot of new
results in the context of braneworld phenomenology compared to standard GR case.
 Most importantly we get a different numerical value of reheating temperature
 as well the extremum temperature compared to the standard GR results.
Next using the extremization principle we justify that the extremum temperature
 is the minimum temperature during reheating which again shows
deviation from standard GR inspired phenomenology. All these facts are reflected
 in the numerical results of the gravitino abundance in reheating and
radiation dominated era. In the context of phase transition
 we also get different numerical results for different
parameters for standard model particle constituents. 

We have further engaged ourselves in investigating for the effect of perturbative reheating.
 To this end  we propose a theory which reflects the effect of particle production
through collision and decay thereby showing a direct connection with the thermalization phenomena.
 To show this link more explicitly we put forward both analytical and numerical expressions for
 the gravitino abundance in a physical volume in the reheating epoch. Next we have found out
 the gravitino abundance in the radiation dominated era.
 Last but not the least we have expressed the total
 gravitino abundance in a final temperature $T_{f}$. Most significantly the precision level of all
 estimated numerical results is the outcome of the $4D$ effective field theory which is analyzes
 with the arrival of lots of sophisticated techniques.

Apart from the aforesaid success in estimating phenomenological
parameters there are some added advantages of our model with reheating in brane which are worth
mentioning. One of the most significant features in the context of braneworld phenomenology is the validity
 of leptogenesis for our model which consequently shows the production of heavy Majorana neutrinos
in the brane.

In future our aim is to search for the signatures
of our model for domain wall formation \cite{mada}
linked to the topological defects, `Q-ball'
formation \cite{kasuya} connected with the
non-topological solitons in braneworld,
the role of Lee-Wick particles in brane reheating and leptogenesis, primordial non-Gaussianity, baryogenesis etc. Last but
not the least the detailed study
 of quantum phase transition using
Monte Carlo simulation technique \cite{liu} to minimize rapid fluctuation  \cite{tye}
 or oscillation during measurement is also an open issue.
We expect to address some of these  issues in near future.


\section*{Acknowledgments}

SC thanks  B. Basu, K. Bhattacharya and A. Mukhopadhyay
for illuminating discussions and Council of Scientific and
Industrial Research, India for financial support through Junior
Research Fellowship (Grant No. 09/093(0132)/2010). SP is supported
by Alexander von Humboldt Foundation, Germany through the project
``Cosmology with Branes and Higher Dimensions'' and is partially
supported by the SFB-Tansregio TR33 ``The Dark Universe''
(Deutsche Forschungsgemeinschaft) and the European Union 7th
network program ``Unification in the LHC era''
(PITN-GA-2009-237920).




\begin{references}

\bibitem{sayan}  S. Choudhury  and  S. Pal,   arXiv:1102.4206 [hep-th].


\bibitem{haw} S. W. Hawking and  G. F. R. Ellis,  Astrophysical Journal, 152  (1968) 25;  S. W. Hawking and  Roger Penrose,  Proceedings of the Royal Society of London,  A  314 (1970) 529.


\bibitem{kofman} L. Kofman,  A. Linde and  A. Starobinsky, Phys. Rev. Lett. 73 (1994) 3195;   L. Kofman,  A. Linde and  A. Starobinsky, Phys. Rev. D 56 (1997) 3258;
   A. Mazumdar and J. Rocher, Phys. Rept. 497  (2011) 85; A. Mazumdar, arXiv:1106.5408 [hep-ph].


\bibitem{burles} S. Burles, K.  M. Nollett and  M. S. Turner, Phys. Rev. D 63 (2001) 063512.



\bibitem{palma}  G. Palma  and  V. H. Cardenas,  Class. Quant. Grav. 18 (2001) 2233;  R. Allahverdi  and  Manuel Drees,  arXiv:hep-ph/0210432;  M. Basler and
B. Kampfer,  International Journal of Modern Physics A 9 (1992)  2033.


\bibitem{Mazumdar} R. Allahverdi, R. Brandenberger, F. Y. C. Racine and A. Mazumdar, Annu. Rev. Nucl. Part. Sci.  60 (2010) 27;  G. Mangano, G. Miele, S. Pastor and M. Peloso,  Phys. Rev. D 64 (2001) 123509;  V.  H. Cardenas, Phys. Rev. D 75 (2007) 083512.


\bibitem{lee} J. H. Traschen and R. H. Brandenberger, Phys. Rev. D 42 (1990) 2491-2504;  A. D. Dolgov and D. P. Kirilova,
Sov. J. Nucl. Phys. 51, (1990) 172-177;  Y. Shtanov, J. H. Traschen and R. H. Brandenberger, Phys. Rev. D 51 (1995) 5438-5455;
 D. Boyanovsky,  M. Dattanasio,  H. J. de Vega,  R. Holman and D. S. Lee, Phys. Rev. D 52 (1995) 6805;
   D. Boyanovsky,   H. J. de Vega and  R. Holman, arXiv:hep-ph/9701304;  J.  McDonald, Phys. Rev. D 61 (2000) 083513;
  D. Boyanovsky,  M. Dattanasio,  H.J. de Vega,  R. Holman,  D. S. Lee  and  A. Singh,  arXiv:hep-ph/9505220.



\bibitem{allah}  R. Allahverdi and  A. Mazumdar,  Phys. Rev. D 76 (2007) 103526;   D. Boyanovsky,  M. Dattanasio,  H.J. de Vega,  R. Holman  and  D. S. Lee,      arXiv:hep-ph/9511361.



\bibitem{seto}  E. J. Copeland  and  O. Seto,  Phys.  Rev. D 72 (2005) 023506;  V.  S. Rychkov  and  A. Strumia,  Phys. Rev. D 75 (2007) 075011.



\bibitem{raga}  R. Rangarajan  and   N. Sahu,  Phys. Rev. D 79 (2009) 103534.



\bibitem{bolz}   M. Bolz, A. Brandenburg  and  W. Buchmuller,  Nucl. Phys. B 606 (2001) 518.



\bibitem{pila}  A. Pilaftsis,  J. Phys. Conf. Ser. 171 (2009) 012017;  S. Davidson, E. Nardi  and Y. Nir,  Phys. Rept. 466 (2008) 105;  W. Buchmuller, R. D. Peccei  and T. Yanagida,  Ann. Rev. Nucl. Part. Sci. 55 (2005) 311;  Y. Nir,  arXiv:hep-ph/0702199;  W. Buchmuller, D. Bari, P. M. Plumacher, Annals Phys. 315 (2005) 305.


\bibitem{van}  J. M. Frere,  F. S. Ling,  M. H. G. Tytgat and  V. V. Elewyck,  Phys. Rev. D 60 (1999) 016005.



\bibitem{yana}  W. Buchmuller, K. Hamaguchi, M. Ibe  and  T. T. Yanagid, Phys. Lett. B 643 (2006) 124;  T. Asaka, K. Hamaguchi and K. Suzuki, Phys. Lett. B 490 (2000) 136-146;  T. Moroi,  H. Murayama  and  M. Yamaguchi, Phys. Lett. B 303 (1993) 289.


\bibitem{Kohri}  K. Kohri, T. Moroi and A. Yotsuyanagi, Phys. Rev. D 73 (2006) 123511;  L. Covi, M. Grefe, A. Ibarra and D. Tran, JCAP 0901 (2009) 029.


\bibitem{Moroi}  T. Moroi, arXiv:hep-ph/9503210.


\bibitem{Choi}  K. Choi, K. Hwang  and H. B. Kim  and T. Lee,  Phys. Lett. B 467 (1999) 211;  R. Rangarajan and N. Sahu, Mod. Phys. Lett. A 23 (2008) 427; R. Allahverdi  and  A. Mazumdar,  JCAP 0610 (2006) 008;  L. Boubekeur, K. Y. Choi, R. R. de Austri and O. Vives,  JCAP 1004 (2010) 005.


\bibitem{prad}  J. Pradler and F. D. Steffen, Phys. Rev. D 75 (2007) 023509.


\bibitem{kolb}  G. F. Giudice,  E. W. Kolb and A. Riotto,  Phys. Rev. D 64 (2001) 023508;  C. Pallis,  Astroparticle Physics  21 (2004) 689-702.



\bibitem{paolo}  P. Creminelli, S. Dubovsky,  A. Nicolis, L. Senatore and  M. Zaldarriaga,  JHEP 0809 (2008) 036;  T. W. B. Kibble, Physics Reports 67  (1980)  183.



\bibitem{chung}  D. J. H. Chung,  E. W. Kolb  and  A. Riotto,  Phys. Rev.  D 60 (1999) 063504;  L . F . Abbott,  E.  Farhi  and  M. B. Wise,  Phys. Lett. 117 B (1982)  29;  J. H. Brodie and D. A. Easson, JCAP 0312 (2003) 004.


\bibitem{turner} E. Kolb and M. Turner , The Early Universe , Addison Wesley (1990).


\bibitem{ali}   A. Linde,  Contemp. Concepts Phys. 5 (2005) 1.


\bibitem{Narlikar} J.V. Narlikar and T. Padmanabhan, Annu. Rev. Astron. Astrophys. 29(1991)
325.


\bibitem{felipe}  R. G. Felipe,  Phys. Lett. B 618 (2005) 7.


\bibitem{ferran}  A. Ferrantelli,   University of Helsinki, Finland, Ph.D. thesis (2010) arXiv:1002.2835 [hep-ph].


\bibitem{stef}   J. Pradler  and  F.  D. Steffen, Phys. Lett. B 648 (2007) 224.


\bibitem{okada}   N. Okada and O. Seto,  Phys. Rev. D 73  (2006) 063505.


\bibitem{mada}  Y. I. Takamizu and K. I. Maeda,  Phys. Rev. D 70 (2004) 123514;  D. Sez and V. J. Ballester, Phys. Rev. D 42 (1990) 3321;  M. F. Parry and A. T. Sornborger,  Phys. Rev. D 60 (1999) 103504.



\bibitem{kasuya}  S. Kasuya  and M. Kawasaki,  Phys. Rev. D 61 (2000) 041301;  T. Hiramatsu, M. Kawasaki and  F. Takahashi,  JCAP 1006 (2010) 008;
M. I. Tsumagari,  Phys. Rev. D 80 (2009) 085010;  R. Allahverdi, A. Mazumdar and  A. Ozpineci,  Phys. Rev. D 65 (2002) 125003.



\bibitem{liu}  J. S. Liu,  Monte Carlo Strategies in Scientific Computing, Springer Publication (2001);
  C. P. Robert and  G. Casella,  Monte Carlo Statistical Methods, Springer Publication (2010); M.Yu.Khlopov, Cosmoparticle physics, World Scientific(1999).



\bibitem{tye}  E. E. Flanagan, S. H. H. Tye  and I. Wasserman,  Phys. Rev. D 62 (2000) 024011.












\end{references}
\end{document}